# Decomposition Formula for the Wall Heat Flux of a Compressible Boundary Layer

Dong Sun[1], Qilong Guo[1], Xianxu Yuan[1,2], Haoyuan Zhang[1,2], Chen Li[1*], Pengxin Liu[1†]

1 State Key Laboratory of Aerodynamics. Mianyang Sichuan 621000, China;

2 Computational Aerodynamics Institute, China Aerodynamics Research and Development Center,

Mianyang Sichuan 621000, China

**Abstract:** Understanding the generation mechanism of the heating flux is essential for the design of hypersonic vehicles. We proposed a novel formula to decompose the heat flux coefficient into the contributions of different terms by integrating the conservative equation of the total energy. The reliability of the formula is well demonstrated by the direct numerical simulation results of a hypersonic transitional boundary layer. Through this formula, the exact process of the energy transport in the boundary layer can be explained and the dominant contributors to the heat flux can be explored, which are beneficial for the prediction of the heat and design of the thermal protection devices.

## 1 Introduction

The heat transfer prediction is of great importance for the hypersonic vehicles. When the transition or turbulence happens, the accurate prediction becomes even more challenging. Therefore, exploring the generation mechanism of the heat flux has drawn wide attention. Many efforts have been put into understanding the generation mechanism of the heat flux, which provides important guidance for the design of the thermal protection system and the thermal management[1].

Some efforts took advantage of the similarities between the generation of friction and heat flux, and constructed the Reynolds analogy[2] $Ra=2St/C_f$ to connect the friction with heat flux, where $St$ is the Stanton number, $C_f$ the skin friction coefficient. Hopkins and Inouye[3] predicted the surface heat flux of the hypersonic boundary layer with $Ra=1$. However, the accuracy of this correlation decreases significantly on a cold wall at high Mach numbers. More intrinsic mechanisms should be considered. Huang et al.[4] derived the formula $q_w = -u_b \tau_w$ by the assumption that the heat transfer into the walls equal the total pressure work done across the channel. Chen et al.[5] established the exact relations for the skin friction with other dynamic and kinetic quantities, and they found these relations revealed that the skin friction is intrinsically coupled with surface temperature through the heat flux. Abe and Antonia[6] proposed a simple relation between the scalar dissipation rate and the wall heat transfer coefficient for the channel flow. Kim et al.[7] proposed a direct approach for the time-dependent heat flux by assuming the temperature approximated as a third-order polynomial of position. Zhang and Xia[8] computed the contributions of the viscous stresses to the heat transfer in a turbulent channel flow by an exact expression, but the turbulent stresses were missing in their expression due to the simplification assumptions used in the channel flow.

Previous studies have not built the relation of the generation mechanism of the heat flux to the dynamic energy transport of the boundary layer. Therefore, more reliable and accurate method to gain insight into the generation mechanism should be derived. In the paper, we proposed a new decomposition formula for the wall heat flux by integrating the conservation equation of total energy, which can explain the energy transport process in the boundary layer and reveal the main factors affecting the wall heat flux.

---

* Corresponding author: lichen@skla.cardc.cn
† Corresponding author: liupengxin@cardc.cn

The work of this paper is organized as follows. In section 2, we describe numerical methods and case setup in brief. In section 3, the detailed derivation of the decomposition formula is presented. In section 4, the validation of the DNS of a hypersonic transitional boundary layer is performed, and the proposed decomposition formula is applied to analyze the heat flux. Finally, some conclusions are drawn in section 5.

**2 Numerical Methods and Case setup**

In order to analyze the heat flux decomposition of the hypersonic transitional boundary layer, a direct numerical simulation is performed. The compressible Navier-Stokes equations in the curvilinear coordinate are adopted as governing equations.

$$\frac{\partial \rho}{\partial t} + \frac{\partial \rho u_j}{\partial x_j} = 0 \quad (1.a)$$

$$\frac{\partial \rho u_i}{\partial t} + \frac{\partial \rho u_i u_j}{\partial x_j} + \frac{\partial p \delta_{ij}}{\partial x_j} = \frac{\partial \sigma_{ij}}{\partial x_j} \quad (1.b)$$

$$\frac{\partial \rho e}{\partial t} + \frac{\partial u_j (\rho e + p)}{\partial x_j} = \frac{\partial u_j \sigma_{ij}}{\partial x_j} + \frac{\partial q_j}{\partial x_j} \quad (1.c)$$

Where $u_1, u_2$ and $u_3$ are streamwise, normal and spanwise velocities, respectively. And $p$ and $\rho$ are pressure and density. The expressions of $\rho e$, $\sigma_{ij}$, and $q_j$ are defined as

$$\rho e = \frac{p}{\gamma - 1} + \frac{1}{2}\rho(u^2 + v^2 + w^2) \quad (2)$$

$$\sigma_{ij} = \frac{\mu}{Re_\infty}\left[\left(\frac{\partial u_i}{\partial x_j} + \frac{\partial u_j}{\partial x_i}\right) - \frac{2}{3}\delta_{ij}\frac{\partial u_k}{\partial x_k}\right] \quad (3)$$

$$q_j = \frac{\mu}{PrMa_\infty^2 Re_\infty (\gamma - 1)}\frac{\partial T}{\partial x_j}. \quad (4)$$

The usual indicial notation is used. Prandtl number $Pr$ is set as 0.72 and the specific heat ratio $\gamma$ is 1.4. The dynamic viscosity $\mu$ is obtained by using Sutherland's law. $Ma_\infty$ is the freestream Mach number and $Re_\infty$ is the Reynolds number. The working fluid is air, and the gas model is assumed to be the perfect gas model.

An in-house code with high-order schemes is employed to perform the DNS. This code has been applied in many DNS simulations of compressible turbulent cases, including compressible homogeneous turbulence[9], turbulent boundary layer[9] and shock wave/boundary layer interaction[10-12]. The accuracy and robustness have been well validated. The hybrid optimized WENO scheme is adopted to discretize the inviscid fluxes with a novel discontinuity detector[13]. In the regions with discontinuities, a seventh-order WENO scheme[14] is activated and in the smooth region, a fourth-order bandwidth-optimized upwind-biased scheme[13] is used to resolve the small structures in the turbulent boundary layer. The viscous terms are discretized by the fourth-order central scheme. The third-order TVD Runge-Kutta method is adopted as the temporal algorithm. $Ma_\infty$ is set as 6.0, and $Re_\infty$ is 12000 which is based on the millimetre and freestream parameters. The freestream temperature is 65K and the wall temperature is 305K. The millimetre is adopted as the reference length.

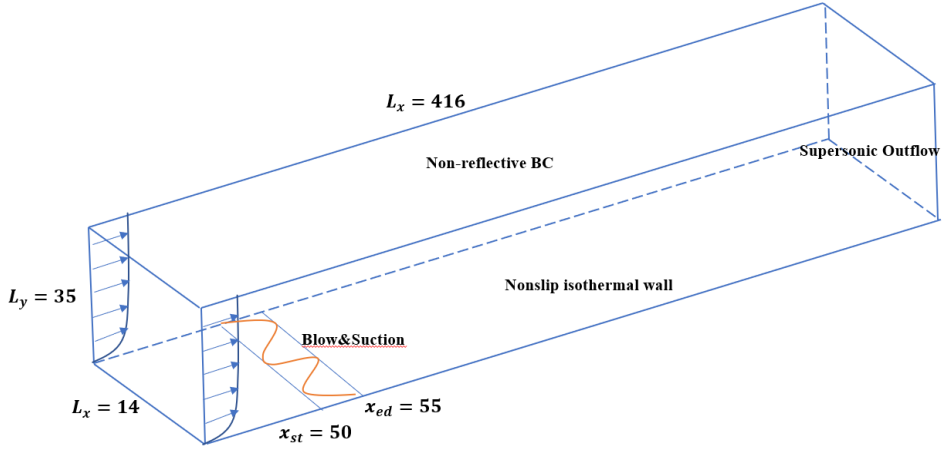

**Figure 1 The sketch of the computational domain**

The sketch of the computational domain is presented in Fig.1. The lengths in the streamwise, normal and spanwise direction are $L_x = 416mm$, $L_y = 35mm$ and $L_z = 14mm$, respectively. The computational domain is discretized with $N_x \times N_y \times N_z = 1151 \times 320 \times 149$ points. The grid points are equally spaced in the spanwise direction, and clustered near the wall in the normal direction. The grid spacing normalized by the wall unit in three directions are $dx^+ = 7.9$, $dy_w^+ = 0.36$ and $dz^+ = 3.4$, respectively. A hypersonic laminar profile is fixed at the inlet. The outlet boundary condition is supersonic outflow with a sponge layer to further suppress the disturbances originated from the outlet. The upper boundary is a non-reflective condition. And an isothermal nonslip condition is applied at the wall. The periodical conditions are set on both sides in the spanwise direction. To trigger a bypass transition, a blow and suction forcing method is employed, which set a normal velocity disturbance $v_n = Af(x)cos(2\pi\lambda_z/L_z)$ at the wall, where $f(x)$ is used to control the streamwise extent of force[15], $\lambda_z$ is the spanwise wavelength and is set as 4.0 in the present simulation.

**Table 1 Expressions for the terms at the right hand of Eq. (5)**

| | | |
|---|---|---|
| Heat conduction | $q_{L,j} = \overline{\kappa\, \partial T/\partial x_j}$ | |
| Turbulent transport of heat | $q_{T,j} = -\overline{\rho u_j'' h''}$ | |
| Molecular diffusion | $D_j = \overline{u_k'' \sigma_{kj}}$ | |
| Turbulent transport of TKE | $T_j = \overline{\rho u_j'' u_k'' u_k''}/2$ | |
| Work by molecular stresses | $MS_j = \tilde{u}_k \overline{\sigma}_{k,j}$ | |
| Work by Reynolds stresses | $RS_j = -\tilde{u}_k \overline{\rho u_k'' u_j''}$ | |

## 3. Derivation of decomposition formula

Under the assumption of homogeneity in the spanwise direction and nonslip condition at the wall, the time-averaged conservative equations for the specific total energy can be expressed as,

$$\bar{\rho}\frac{D\tilde{E}}{Dt} + \frac{\partial \bar{p}\tilde{u}_j}{\partial x_j} = \frac{\partial}{\partial x_j}(q_{L,j} + q_{T,j} + D_j + T_j + MS_j + RS_j) \tag{5}$$

when $\phi$ is an arbitrary variable, $\bar{\phi}$ denotes Reynolds average, $\tilde{\phi}$ Favre average and $\phi''$ is the fluctuations concerning the Favre average. $\tilde{E}$ denotes specific total energy, $q_{L,j}$ the heat conduction, $q_{T,j}$ the turbulent transport of heat, $D_j$ the molecular diffusion, $T_j$ the turbulent transport of turbulent kinetic energy (TKE), $MS_j$ work by the molecular stresses and $RS_j$ work by Reynold stresses. In Table 1, the specific expressions of the terms at the right hand of Eq. (5) are presented. More details about Eq. (5) can be found in the book of Wilcox[16].

The work of Renard[17] and Li[18] about skin friction decomposition provides good hints for building the integration relation of heat flux. Transformed the initial frame into a reference absolute one by assuming that the wall moves at the speed $U_\infty$. The expressions of the time $t_a$, coordinates $x_a$, $y_a$, velocities $u_a$, $v_a$, pressure $p_a$ and the density $\rho_a$ in the reference absolute frame satisfy

$$t_a = t, x_a = x - U_\infty t, y_a = y,$$
$$u_a = u - U_\infty, v_a = v, \tag{6}$$
$$p_a = p, \quad \rho_a = \rho$$

where the subscript '$a$' represents 'absolute' variables under the reference frame. By substituting Eq. (6) into Eq. (5), Eq. (5) takes the form,

$$\bar{\rho}_a \frac{D\tilde{E}_a}{Dt_a} + \frac{\partial(\bar{p}_a \tilde{u}_a)}{\partial(x_a + U_\infty t_a)} + \frac{\partial(\bar{p}_a \tilde{v}_a)}{\partial y_a}$$
$$= \frac{\partial}{\partial y_a}(q_{L,y,a} + q_{T,y,a} + D_{y,a} + T_{y,a} + MS_{y,a} + RS_{y,a}) \tag{7}$$
$$+ \frac{\partial}{\partial(x_a + U_\infty t_a)}(q_{L,x,a} + q_{T,x,a} + D_{x,a} + T_{x,a} + MS_{x,a} + RS_{x,a})$$

By multiplying $\tilde{u}_a$ at both sides of Eq. (7), combining the continuity equation and moving the heat fluxes in the normal direction to the left side and all the other terms to the right, the following equation could be obtained,

$$\tilde{u}_a \frac{\partial}{\partial y_a}(q_{L,y,a} + q_{T,y,a})$$
$$= \tilde{u}_a \bar{\rho}_a \frac{D\tilde{E}_a}{Dt_a} + \tilde{u}_a \frac{\partial(\bar{p}_a \tilde{u}_a)}{\partial(x_a + U_\infty t_a)} + \tilde{u}_a \frac{\partial(\bar{p}_a \tilde{v}_a)}{\partial y_a}$$
$$- \tilde{u}_a \frac{\partial}{\partial y_a}(D_{y,a} + T_{y,a} + MS_{y,a} + RS_{y,a}) \tag{8}$$
$$- \tilde{u}_a \frac{\partial}{\partial(x_a + U_\infty t_a)}(q_{L,x,a} + q_{T,x,a} + D_{x,a} + T_{x,a} + MS_{x,a} + RS_{x,a})$$

Integrating Eq. (8) from the wall to the farfield, the heat flux on the wall ($y$=0) can be expressed as the following form

$$q_{L,y,a}|_{y=0} =$$
$$\frac{1}{U_\infty} \int_0^\infty \left[(q_{L,y,a} + q_{T,y,a})\frac{\partial \tilde{u}_a}{\partial y_a}\right] dy$$
$$- \frac{1}{U_\infty} \int_0^\infty \left[\tilde{u}_a \frac{\partial}{\partial y_a}(D_{y,a} + T_{y,a} + MS_{y,a} + RS_{y,a})\right] dy \tag{9}$$
$$+ \frac{1}{U_\infty} \int_0^\infty \left[\tilde{u}_a \bar{\rho}_a \frac{D\tilde{E}_a}{Dt_a} + \tilde{u}_a \left(\frac{\partial \bar{p}_a \tilde{u}_a}{\partial(x_a + U_\infty t_a)}\right) + \tilde{u}_a \frac{\partial(\bar{p}_a \tilde{v}_a)}{\partial y_a}\right] dy$$
$$- \frac{1}{U_\infty} \int_0^\infty \left[\tilde{u}_a \frac{\partial}{\partial(x_a + U_\infty t_a)}(q_{L,x,a} + q_{T,x,a} + D_{x,a} + T_{x,a} + MS_{x,a} + RS_{x,a})\right] dy$$

Finally, Eq. (9) can be transformed into the initial frame and the heat flux coefficient are obtained by introducing the definition of the coefficient $C_h = \frac{1}{\rho_\infty U_\infty^3} \kappa \frac{\partial T}{\partial y}\big|_w$,

$$C_h = \underbrace{\frac{1}{\rho_\infty U_\infty^4} \int_0^\infty q_{L,y} \frac{\partial \tilde{u}}{\partial y} dy}_{C_{h,1}} + \underbrace{\frac{1}{\rho_\infty U_\infty^4} \int_0^\infty q_{T,y} \frac{\partial \tilde{u}}{\partial y} dy}_{C_{h,2}} - \underbrace{\frac{1}{\rho_\infty U_\infty^4} \int_0^\infty (\tilde{u} - U_\infty) \frac{\partial D_y}{\partial y} dy}_{C_{h,3}}$$
$$- \underbrace{\frac{1}{\rho_\infty U_\infty^4} \int_0^\infty (\tilde{u} - U_\infty) \frac{\partial T_y}{\partial y} dy}_{C_{h,4}} - \underbrace{\frac{1}{\rho_\infty U_\infty^4} \int_0^\infty (\tilde{u} - U_\infty) \frac{\partial MS_y}{\partial y} dy}_{C_{h,5}} - \underbrace{\frac{1}{\rho_\infty U_\infty^4} \int_0^\infty (\tilde{u} - U_\infty) \frac{\partial RS_y}{\partial y} dy}_{C_{h,6}} \quad (10)$$
$$+ \underbrace{\frac{1}{\rho_\infty U_\infty^4} \int_0^\infty (\tilde{u} - U_\infty) \left[ \bar{\rho} \frac{\partial \tilde{E}}{\partial t} + \frac{\partial \bar{p}\tilde{u}}{\partial x} + \frac{\partial \bar{p}\tilde{v}}{\partial y} - \frac{\partial}{\partial x}(q_{L,x} + q_{T,x} + D_x + T_x + MS_x + RS_x) \right] dy}_{C_{h,7}}$$

In Eq. (10), the heat flux coefficient $C_h$ has been decomposed into seven parts, i.e., $C_{h,1}$ to $C_{h,7}$. The first part $C_{h,1}$ represents the contribution of heat conduction, the second part $C_{h,2}$ the turbulent transport of heat, the third part $C_{h,3}$ the normal component of the molecular diffusion, the fourth part $C_{h,4}$ the turbulent transport of turbulent kinetic energy (TKE), the fifth part $C_{h,5}$ the work by the molecular stresses and the sixth part $C_{h,6}$ the work by Reynolds stresses. And the last part $C_{h,7}$ contains the streamwise heterogeneity, pressure work and the variation of the specific total energy with time.

**4 Decomposition Results of Wall Heat Flux**

In this section, the reliability and accuracy of the new decomposition formula will be demonstrated by the DNS results of a hypersonic transitional boundary layer. The contributions of different energy transport processes such as the heat conduction, the turbulent transport of heat, the work done by molecular streses will be calculated. Moreover, the key normal locations where the structures impacting the wall heat will be assessed by the integrand functions of these contributions.

**4.1 DNS results of the hypersonic transitional boundary layer**

The instantaneous vortical structures displayed by $Q$ criterion[19] is presented in Fig. 2. The transition process can be observed. Three streamwise locations (P1, P2 and P3) are chosen to validate the decomposition results, which are marked by shade planes. In Fig. 3, the mean streamwise velocity after van Driest transformation at P3 is presented and compared with DNS results of hypersonic results by Priebe[20]. Good agreement is obtained. The discrepancy in the outer part of the boundary layer is caused by the different friction velocity at the wall. In Fig. 4, the Reynolds stresses normalized by friction velocity are presented. Several supersonic and hypersonic results[21-23] are also displayed. The present data generally agree with those in the reference results. These comparisons demonstrate that the performance of our present direct simulation is acceptable, and the obtained data can be used for decomposition analysis.

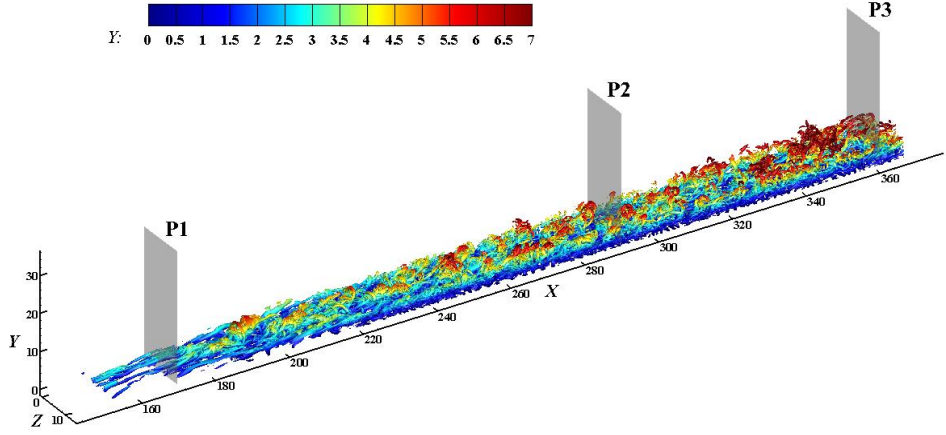

**Fig. 2 The instantaneous magnitude of the vortical structures**

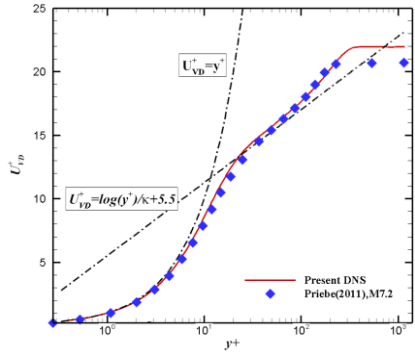

Fig. 3 van Driest transformed mean streamwise velocity

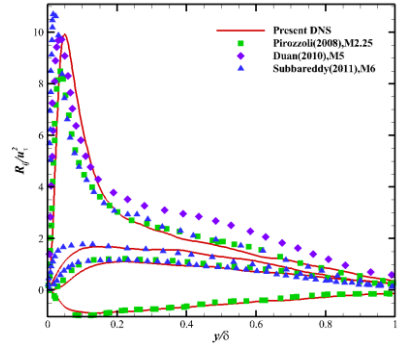

Fig. 4 Streamwise turbulent intensity normalized by Morkovin transformed velocity

Table 2 Contributions of the terms in Eq. (10)

|    | $C_{h,1}/C_{h,decom}$ | $C_{h,2}/C_{h,decom}$ | $C_{h,3}/C_{h,decom}$ | $C_{h,4}/C_{h,decom}$ | $C_{h,5}/C_{h,decom}$ | $C_{h,6}/C_{h,decom}$ | $C_{h,7}/C_{h,decom}$ |
|----|------|------|------|------|------|------|------|
| P1 | -0.111 | -1.379 | 0.018 | -0.084 | 0.478 | 2.437 | -0.359 |
| P2 | -0.213 | -0.785 | 0.012 | -0.072 | 0.673 | 1.299 | 0.086 |
| P3 | -0.208 | -0.713 | 0.015 | -0.054 | 0.662 | 1.200 | 0.098 |

**4.2 Contributions of different energy transport processes**

The contribution of the different energy transport process to the wall heat flux at different streamwise locations, e.g. P1-P3 are calculated according to Eq. (10). The ratios of the contributions of the terms in Eq. (10) to the reconstructed heat flux coefficient $C_{h,decom}$ is presented in Table 2 and the relative errors are presented in Table 3. The relative error is defined as $Error = (C_{h,decom} - C_{h,0})/C_{h,0} \times 100\%$, where $C_{h,0}$ is the time-averaged heat flux coefficient. The relative errors are very small in both transitional and turbulent regions. In Table 2, it is found that the work done by the molecular stresses ($C_{h,5}$) and Reynolds stresses ($C_{h,6}$) play dominant roles in heat production. The sum of the contributions of these two terms takes over 2.7 times of $C_{h,0}$ in the transition region and 1.9 times in the turbulent region. The contributions of the heat conduction ($C_{h,1}$) and the turbulent transport of heat ($C_{h,2}$) are negative, which indicates both processes act as the transporters of the heat and carry the extra heat to the outer regions of the boundary layer.

**Table 3 Relative errors between $C_{h,decom}$ and $C_{h,0}$ at P1 to P3**

|  | P1 | P2 | P3 |
|---|---|---|---|
| $C_{h,0}$ | 1.195E-4 | 1.825E-4 | 1.616E-4 |
| $C_{h,decom}$ | 1.183E-4 | 1.819E-4 | 1.620E-4 |
| Relative Error | 1.00% | 0.33% | 0.25% |

The trends of the contributions can be assessed when more sample locations are considered. The trends of the contributions in the transition process are presented in Fig. 5. In addition, the instantaneous the density gradient and the time-averaged heat flux coefficient are also presented to show the overall evolution procedure. The symbols in the line of $C_{h,i}$ indicate different sample locations (from left to right labelled as X0 to X10). The first sample point (X0) located in the laminar region; therefore, the work of the molecular stress and the heat conduction is the main factors affecting the wall heat. After entering the transition region, the effects of the Reynold stresses become dominant. Moreover, a rapid variation of $C_{h,6}$ is observed in the transition region due to the nonlinear development of the coherent structures. While in the turbulent region, an equilibrium state is reached and a smooth variation is obtained. The trends of contributions of the heat conduction $C_{h,1}$ and the molecular stresses $C_{h,5}$ are very smooth in both transition and turbulent regions which indicate they are not sensitive to the change of the turbulent fluctuations. In addition, the effects of the molecular diffusion ($C_{h,3}$), the turbulent transport of TKE ($C_{h,4}$) and the streamwise heterogeneity ($C_{h,7}$) are very small during the whole transition.

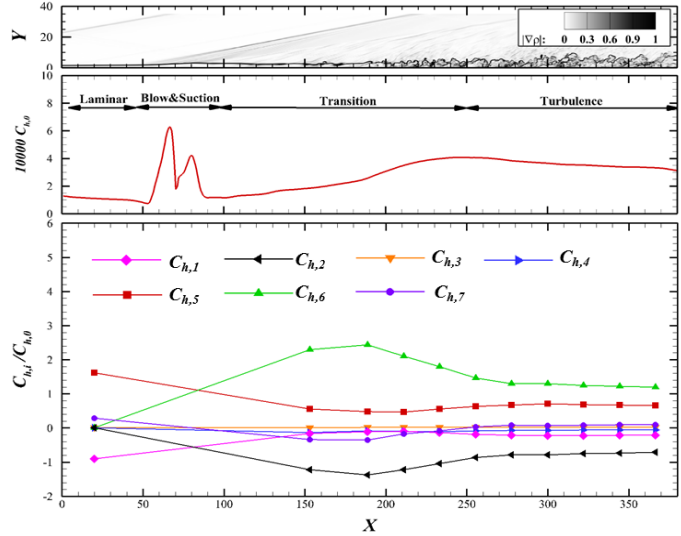

**Fig. 5 the streamwise trends of the contributions of the terms of heat flux decomposition**

### 4.3 Key normal locations affecting the wall heat

The key normal locations where the structures affecting the wall heat can be attained by analyzing the integrand functions of the terms in Eq. (10). Two main transporters, i.e., the heat conduction ($C_{h,2}$) and the turbulent transport ($C_{h,3}$) and two main contributors, i.e., the work by molecular stresses ($C_{h,5}$) and Reynolds stresses ($C_{h,6}$) are analyzed in this section.

The integrand functions of the heat conduction ($C_{h,2}$) and the turbulent transport ($C_{h,3}$) are presented in Fig. 6. The lines in the figure denote the locations in the transitional region and the symbols are the locations in the turbulent region. The contribution of $C_{h,2}$ is positive when the normal locations are near the wall due to the positive temperature gradient. When $y^+$ reaches 10, a valley is observed, which is formed by the decrease of the temperature after the peak of the mean temperature. Compared with the integrand function of $C_{h,2}$, two valleys are observed in the integrand functions of $C_{h,3}$. The first valley locates at $y^+ \approx 65$, the magnitude of the valley reduces and finally disappears as the locations move downstream. The second valley locates around $y^+ = 10$, which corresponds to the buffer layer of the turbulent boundary layer. The valley will get larger as the locations move downstream.

The key normal locations of the work by molecular stresses ($C_{h,5}$) and Reynolds stresses ($C_{h,6}$) are analyzed in Fig. 7. The positive contribution of $C_{h,5}$ mainly located in the regions near the wall. As the normal location increases, the integrand function will decrease and a valley is formed around $y^+ = 10$. Meanwhile, a peak is observed around $y^+ = 10$ in the integrand function of contribution of $C_{h,6}$. The peak locates in the buffer layer of the boundary layer. In addition, in the transition process, the magnitude of the peak will get larger as the streamwise distance increases.

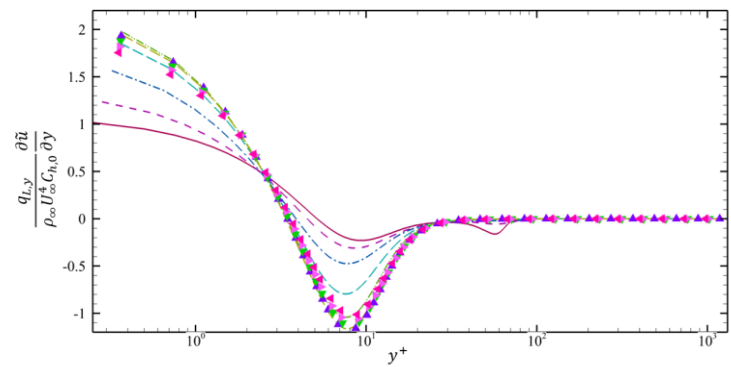

(a)

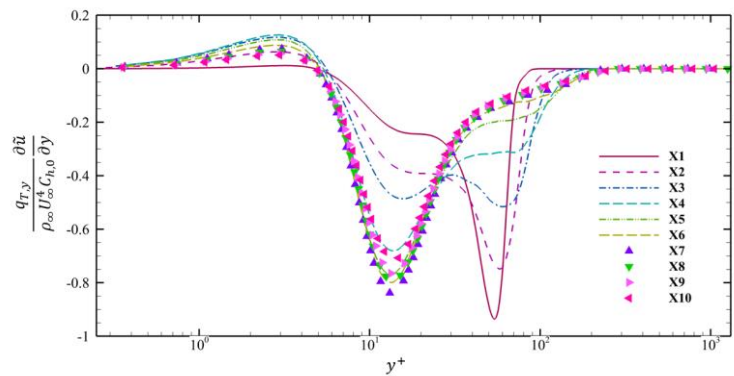

(b)

**Fig. 6 The distribution of the integrand functions of the contributions of the heat conduction $C_{h,2}$ (a) and the turbulent transport $C_{h,3}$ (b).**

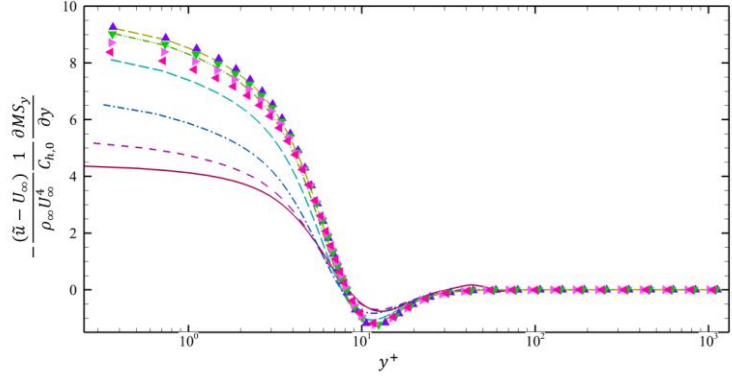

(a)

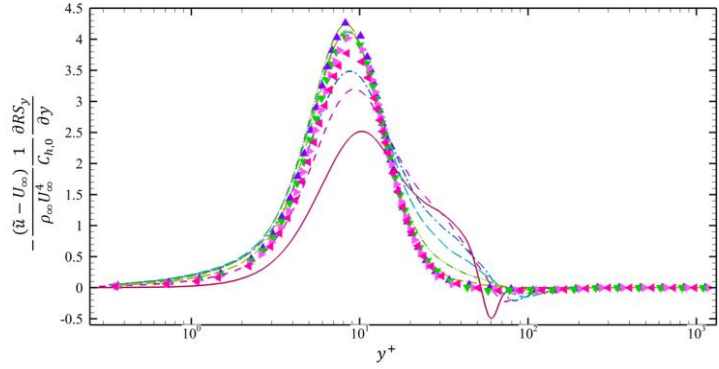

(b)

**Fig. 7 The distribution of the integrand of the contribution of work done by molecular stresses $C_{h,5}$ (a) and Reynolds stresses $C_{h,6}$ (b)**

**5 Conclusions**

In this paper, we proposed a new decomposition formula for the wall heat flux. And the performance of the formula has been well demonstrated by DNS results of a hypersonic transitional boundary layer.

(1) Through this formula, the wall heat flux can be decomposed into contributions of seven terms, i.e., the heat conduction, turbulent transport of heat, molecular diffusion, turbulent transport of TKE, molecular stresses, the Reynolds stresses and the streamwise heterogeneity.

(2) The contributions of each term can be calculated quantitatively. For the present case, it is found that the heat flux produced by the work done by Reynolds stresses and molecular stresses is much higher than the time-averaged heat flux on the wall. The heat conduction and turbulent convection will carry the extra heat into the outer part of the boundary layer.

(3) The structures in the buffer layer ($y^+ = 10$) play a dominant role in the production and transfer of the heat flux for the present simulation. Moreover, the contributions of the heat conduction and the molecular stresses are mainly affected by the gradient of the temperature.

As the assumption used in the derivation is only the spanwise homogeneity and nonslip wall, this formula can be applied in the analysis of the heat transfer of a hypersonic transitional/turbulent boundary layer at high Mach numbers, which can be employed to identify the main factors affecting the wall heating and provide good guidances in the design the thermal protection system.

**Acknowledgements**

NA.

## Authors' contributions

The major contribution of this manuscript is due to the first author, while the second author gives important suggestion on the derivation of the formulas. The author(s) read and approved the final manuscript.

## Funding

This work was supported by the National Key Research and Development Program of China (Grant No. 2019YFA0405201), the National Natural Science Foundation of China (Grant No.11802324), and the National Numerical Windtunnel Project.

## Availability of data and materials

The data that support the findings of this study are available from the corresponding author upon reasonable request.

## Competing interests

The authors declare that they have no competing interests